\documentclass[useAMS,usenatbib]{mn2e}
\usepackage{epsfig}
\usepackage{graphicx}
\usepackage{amssymb}
\usepackage{epstopdf}
\title{Limits on orbit crossing planetesimals in the resonant multiple planet system, KOI-730}              
\author[]{Alexander Moore \& Imran Hasan, \& Alice C. Quillen \\
{Department of Physics and Astronomy, University of Rochester, Rochester, NY 14627, USA
}  \\
}
\begin{document}

\maketitle
\begin{abstract}
A fraction of multiple planet candidate systems discovered from transits by the Kepler mission contain
pairs of planet candidates that are in orbital resonance or are spaced slightly too far apart to be in
resonance. We focus here on the four planet system, KOI 730, that has planet periods satisfying the
ratios 8:6:4:3. By numerically integrating four planets initially in this resonant configuration in
proximity to an initially exterior cold planetesimal disk, we find that of the order of a Mars
mass of planet-orbit-crossing planetesimals is sufficient to pull this system out of resonance.
Approximately one Earth mass of planet-orbit-crossing planetesimals increases the interplanetary spacings
sufficiently to resemble the multiple planet candidate Kepler systems that lie just outside of resonance.
This suggests that the closely spaced multiple planet Kepler systems, host only low mass debris disks or
their debris disks have been extremely stable. We find that the planetary inclinations increase as a
function of the mass in planetesimals that have crossed the orbits of the planets. If systems are left at
zero inclination and in resonant chains after depletion of the gas disk then we would expect a
correlation between distance to resonance and mutual planetary inclinations. This may make it possible
to differentiate between dynamical mechanisms that account for the fraction of multiple planet systems
just outside of resonance.
%Planetary systems comprised of closely spaced low mass planets (sub Neptune) could never have
%experienced an event involving disruption of a 50 Earth mass planetesimal belt as in our solar system.
\end{abstract}

\section{Introduction}
The latest tally of multiple planet candidate systems discovered by the Kepler mission \citep{batalha12}
includes 361 multiple-planet systems \citep{fabrycky12}. A statistical analysis, focused on the
probability that binary stars are the most likely contaminant, finds that most of the multiple planet
candidates are real planetary systems \citep{liss12}. The large number of recently discovered multiple
planet systems represents a significant (by an order of magnitude) increase in the number of
known multiple planet systems compared to those discovered from radial velocity surveys \citep{wright11}. 

Both transit and radial velocity discovered multiple planet systems contain pairs of planets that are
in first order mean motion resonance \citep{wright09,liss11,wright11}. In the Kepler transit systems,
there are statistically significant excesses of candidate planet pairs both in resonance and spaced
slightly too far apart to be in resonance (as delineated by \citealt{veras12}), particularly near the
2:1 mean motion resonance \citep{liss11,fabrycky12}.

Planet migration due to tidal interaction with a gas disk is a possible mechanism through which
convergent migration induces resonance capture, leaving planets in resonance
(e.g. \citealt{lee02,ferraz03,kley04,lee04,papa05,thommes08,morbi07,libert11,rein12}).
Gravitational interactions between planets leading to planet-planet scattering events
(e.g., \citealt{god09,fabrycky10,moore12,rein12}), turbulence in the disk \citep{pierens11},
and tidal interactions between planets \citep{papa11,lithwick12} have been proposed as mechanisms for
pulling initially resonant planetary systems out of resonance. Interactions with planetesimals, for
example as part of the `Nice' model for the early solar system evolution, can also cause planets to
diverge away from resonance \citep{nice} (also see \citealt{thommes08}).

Simulations of planets and planetary embryos embedded in a gas disk allow planets to become trapped in
resonance \citep{kley04,morbi07,rein12}. After the gas disk dissipates, newly formed planets may be left
in a chain of mean-motion resonances \citep{morbi07,matsumura10,moeckel12}.  By a chain of mean motion
resonances, we mean that each consecutive pair of planets is in a $j+1:j$, first order, mean motion
resonance (though the integer $j$ can differ for each pair). Resonant chains have been chosen as initial
conditions for studies of planetary system evolution after the depletion of the gas disk (e.g.,
\citealt{nice,thommes08,batygin10}).
%consequently it may be that when a gas disk dissipates, planets are left in a chain of mean motion resonances.
When pairs of planets are in or near mean motion resonances, there may be a librating or nearly fixed
Laplace angle involving three or more bodies. Three-body resonances (e.g., those comprised of zero-th
order terms; \citealt{quillen11}) may be present even when pairs of planets are not in first order mean
motion resonances. A system in a resonant chain of first order resonances does not necessarily exhibit
a librating Laplace angle involving three or more bodies.

The four planet candidate system,  KOI-730, contains planets with periods that satisfy the ratios 8:6:4:3 
to approximately 1 part in 1000 or better \citep{liss11}.  These ratios imply that the outer two and
inner two planet pairs are in (or near) a 4:3 mean motion resonance and that the middle pair of planets is in
(or near) a 3:2 resonance. In this study we explore the evolution of a similar model four planet system
in proximity to a planetesimal disk. We ask: how much mass in orbit-crossing planetesimals is sufficient
to pull this system out of resonance?

For a pair of planets and a first order mean motion resonance, \citet{fabrycky12} define a parameter, 
$\zeta_{1,1}$,  that measures proximity to a first order mean motion resonance,
\begin{equation}
\zeta_{1,1} \equiv 2 \left( {1 \over {\mathcal P} - 1} - {\rm Round}\left({1 \over {\mathcal P } - 1} \right) \right), \label{eqn:zeta}
\end{equation}
where ${\mathcal P} = P_i/P_j$ (greater than 1) is the observed period ratio of planets $i,j$ (and as defined in the appendix by \citealt{fabrycky12}). 
%
%$\footnote{The $\zeta_1$ parameter adopted by \citet{liss11} actually varies from -1.5 to 1.5 but differs only by that used here by a factor of 1.5, see the appendix by \citet{fabrycky12} for  the  definition involving two indices}.
The `Round' function rounds to the nearest integer. The parameter $\zeta_{1,1} = 0$ at a $j+1:j$ first
order resonance and the function goes from -1 to 1 at second order resonances ($j+2:j$). The parameter
used by \citet{liss11}, $\zeta_1 = 1.5\zeta_{1,1}$ and varies from -1.5 to 1.5.

The probability density distribution of $\zeta_{1}$ values generated from all pairs of Kepler planet
transit candidates, for pairs residing in a single system, exhibits a peak at about
$\zeta_{1} \approx - 0.2$, (see Figure 11 by \citealt{liss11} and  Figure 5 by \citealt{fabrycky12}).
For KOI-730, $\zeta_1 = -0.0123, -0.0186, -0.0063$ for the inner pair, middle pair and outer pair of
planets, respectively \citep{fabrycky12}, consequently the system is likely to be in or very near a
resonant chain. By integrating a system modeled after the KOI-730 system that is initially in a chain
of resonances, we ask: how much mass in orbit-crossing planetesimals would increase the interplanetary
spacing so that $\zeta_1 \sim -0.2$? We also note that due relatively low mass of the planets and their
proximity to the star, their resonant widths are likely smaller than $\zeta_1 = 0.1$, as first order
mean motion resonance width scales approximately with mass to the 2/3 power (e.g. \citet{wisdom80}).

We first describe how we find resonant chain configurations for the KOI-730 system.  We use these
configurations to construct initial conditions for N-body integrations that include a planetesimal disk.   
A summary and discussion follows.

\begin{table}
\caption{\large Masses and Periods for the planet candidates in the KOI 730 system 
and initial orbital elements \label{tab:tab1}}
\begin{tabular}{@{}lccccc}
\hline
Mass 			& Period	& $a$		& $e$		& $\omega$	& $M$		\\
($M_\oplus$) 	& (days)	&			&			&			&			\\
2.5				& 7.3840	& 1.0		& 0.055589	& -2.808830	&  1.216544	\\
3.7				& 9.8487	& 1.211221	& 0.071128	& -0.739956	&  2.841707	\\
8.6				& 14.7884	& 1.586171	& 0.050110	&  1.453201	& -0.924252	\\
6.2				& 19.7213	& 1.920630	& 0.043428	& -1.990222	& -0.172501	\\
%7.5030e-06 1.000000 0.055589 0.000000 0.000000 -2.808830 1.216544  #m a e i longnode argperi meananom
%1.1100e-05 1.211221 0.071128 0.000000 0.000000 -0.739956 2.841707  #m a e i longnode argperi meananom
%2.5960e-05 1.586171 0.050110 0.000000 0.000000 1.453201 -0.924252  #m a e i longnode argperi meananom
%1.8690e-05 1.920630 0.043428 0.000000 0.000000 -1.990222 -0.172501 #m a e i longnode argperi meananom 
\hline
\end{tabular}
{
Planet masses for the KOI 730 system are computed from radii based on transit durations taken from
http://archive.stsci.edu/kepler/planet\_candidates.html \citep{batalha12}, using equation \ref{eqn:mass}.
The observed periods are given in days and are taken from the same website. The rightmost four columns
give the orbital elements we used as initial conditions (see section 2). Here $\omega$ is the argument
of pericenter and $M$ the mean anomali. Initial inclinations and the longitudes of the ascending node
were set to zero. These orbital elements were taken from the integration shown in Figure \ref{fig:imran}.
}
\end{table}

\begin{table}
\caption{\large N-body Simulations \label{tab:tab2}}
\begin{tabular}{@{}lcc}
\hline
 Simulation	& Planetesimal Mass		& Orbit-crossing mass	\\
			& ($M_\oplus$)			& ($M_\oplus$)			\\
 Z			& 0						& 0						\\
 M			& $10^{-4}$				& 0.04					\\
 E3			& $3.3 \times 10^{-4}$  & 0.12					\\
 E			& $10^{-3}$				& 0.46					\\
 N5			& $3.3 \times 10^{-3}$	& 1.7					\\
 N			& $1.67 \times10^{-2}$	& 16.6					\\
\hline
\end{tabular}
{\\
Each simulated disk contains 1024 equal mass planetesimals. The planetesimal masses in units of Earth
mass are listed in the second column. The third column shows the total mass in planetesimals (in Earth
masses) that crossed the orbits of any of the planets at the end of the simulation. The names of the
simulations are related to the masses of the planetesimal disks. The Z simulation has a zero mass disk.
The M, E and N simulations have disks with Mars, Earth and Neptune masses, respectively.
The E3 and N5 simulations have disks with a third Earth and a fifth Neptune mass, respectively.
}
\end{table}

\section{Setting up resonant chain configurations}
We first describe how we find orbital elements consistent with a 8:6:4:3 ratio resonant chain for a four
planet system similar to the KOI-730 multiple planet system.   
% \citet{liss11} report periods and radii for the four planet candidates in the KOI-730 system.  To
%numerically integrate the system, we must chose initially values for all six orbital elements for each planet.
We find resonant chain configurations by integrating a 5 body system under the influence of gravity
(four planets and the central star) and including a Stokes drag-like form for dissipation that induces
both migration and eccentricity damping (as previously done by \citealt{beauge06,batygin10,libert11}).
The drag gives a force per unit mass in the form adopted by \citet{beauge06}
\begin{equation}
%{d^2 {\bf r} \over dt^2} 
{\bf F}_{drag} = - {{\bf v} \over 2 \tau_a} - {{\bf v - v}_c \over \tau_e}
\end{equation}
where $\bf v$ is the planet velocity and ${\bf v}_c$ is the velocity of a planet in a circular orbit
at the current radius (from the star) of the planet. We use a 4th order adaptive step-size Hermite
integrator (that described by \citealt{makino92}) with the addition of the above drag force. We work in
units of the innermost planet's initial orbital period or 7.3840 days.

The drag force induces radial migration on a timescale ${a \over \dot a} \sim \tau_a$, and eccentricity
damping on a timescale ${e \over \dot e} \sim \tau_e$ where $a,e$ are the planet's semi-major axis and
eccentricity, respectively. Resonant capture can only occur when the drift rate, $\dot a$, is
sufficiently slow such that the square of libration frequency in resonance exceeds the drift rate (this
defines the adiabatic limit; \citealt{quillen06}). Following resonant capture, eccentricities of planets
can increase as they drift inwards, \citep{lee02}, causing instability (e.g.,
\citealt{kley04,libert11,rein12}). For two planet systems, an equilibrium state may be reached that
depends on the ratio $K \equiv {\tau_a \over \tau_e}$  \citep{lee02}. As discussed by \citet{rein12},
if two bodies lie initially outside the 2:1 or 3:2 resonance, then they are more likely to capture in
one of those than in the 4:3 resonance. Here we require that the outer and inner pair of planets are
captured into the 4:3 resonance, consequently we began the integration with planets spacings just
outside the desired resonances. The migration rates were adjusted so that the migration is sufficiently
fast that capture into weaker second order resonances such as the 5:3 or the 7:5 is unlikely.

The masses of the four planets were set from the radii measured from the transit durations and reported
by \citet{batalha12}. We adopt the power-law relationship for planetary mass, $M_p$, as a function of
radius, $R_p$, used by \citet{liss11,fabrycky12}, 
\begin{equation}
M_p=M_\oplus (R_p /R_\oplus)^\alpha,  \label{eqn:mass}
\end{equation}
where $M_\oplus, R_\oplus$ are the mass and radius of the Earth, and the exponent $\alpha  = 2.06$ for
$R_p > R_\oplus$. %and ? = 3 for Rp  R?.
This choice is motivated by Solar System planets as it is a good fit to Earth, Uranus, Neptune, and
Saturn. KOI-730's surface gravity and effective temperature (reported in Table 9 by \citealt{batalha12},
with surface gravity log${_10}$ g(in cgs) = 4.39 and effective temperature $T_{eff}$ = 5599 K) are
similar to that of the Sun (with log$_{10}$=4.43 and $T_{eff}=5780$K) so we computed planet to stellar
mass ratios from the estimated planet radii, equation \ref{eqn:mass} and using a Solar mass for the host
star. Estimated planet masses and their periods are summarized in Table \ref{tab:tab1}.
The estimated planet masses for the KOI-730 system are much lower than the approximately Jupiter mass
planets considered by \citet{rein12} for the HD2006964 system. The 4:3 resonance may be more stable
in lower mass systems. 
 
%The log10 of the surface gravity of the Sun (in cgs) is 4.43, wheras that for KOI-730 is 4.39.
%The effective temperature of the Sun is 5780 K whereas that for KOI-730 is 5599 K.

\begin{figure}
\includegraphics[width=9cm]{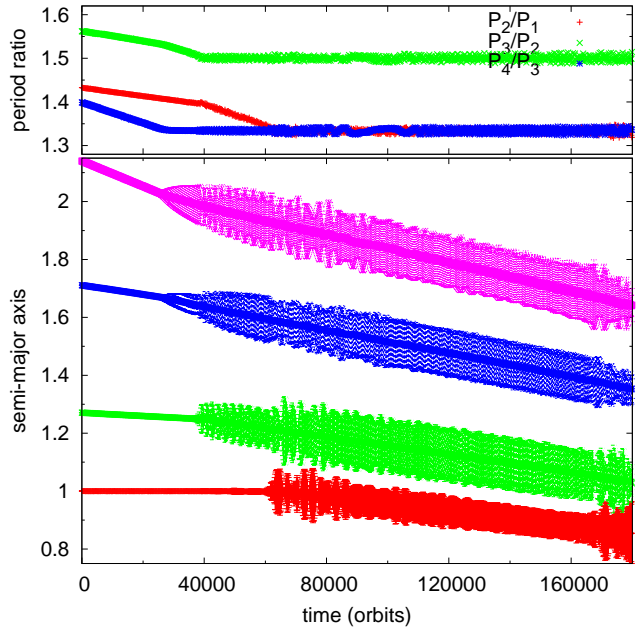}
\caption{a)Semi-major axes of the four planets under the influence of a forced dissipative process as a
function of time.
b) Period ratios for the same integration. This integration was used to generate initial orbital elements
for the four planets in the simulations including planetesimals. We took orbital elements from the time
at $70,000$ orbits.
}
\label{fig:imran}
\end{figure}

An integration is shown in Figure \ref{fig:imran} with a) showing the four semi-major axes as a function
of time and b) showing the period ratios of consecutive pairs of planets. Semi-major axes are shown in
units of the initial innermost planet's semi-major axis.  The timescales $\tau_a$ and $\tau_e$ can be 
chosen separately for each planet. We set $\tau_a$ for the innermost planet to be extremely long ($10^9$
orbits), and that for the outer 3 planets to be progressively shorter ranging from $10^6$ to $10^5$
orbits. Note that $10^5$  orbits of the innermost planet is only approximately 2000 years. We arranged
the drift rates so that the outer planet migrates more quickly than the inner ones. After the outer
planet captures the third planet, the two together migrate more slowly than the outer planet. To maintain
a constant drift rate for the pair, the third planet was set with a longer value for $\tau_a$, and
similarly, $\tau_a$ for the second planet was chosen to be larger than that for the third planet.
  
Once in resonance, the eccentricities of the planets increase as the planets drift inwards. A steady
state can be reached with eccentricity values that depend on the size of the eccentricity damping, set
here with $\tau_e$ \citep{lee04}. High values of $\tau_e$ (corresponding to low levels of damping) are
associated with high planet eccentricities, whereas low values of $\tau_e$ reduce the eccentricities of
the planets. We found that this system became unstable without significant eccentricity damping.
Consequently we set the level of eccentricity damping sufficiently high, $\tau_e \sim 10^4$, so that the
system remained in resonance after all planets were captured into resonance. This value is high compared
to that predicted for tidal interactions between disk and planet. Large $K$ values (up to 100) have been
used previously \citep{batygin10} to generate stable initial resonant conditions for subsequent
integration, as we do here. As can be seen from Figure \ref{fig:imran} the outer pair of planets first
captures into resonance, then the middle pair and lastly the innermost pair is captured into resonance.
\citet{morbi07} stressed that the order of captures can affect the resonant angles.
The 8:6:4:3 ratios imply that the second and fourth planets are near or in a 2:1 mean motion resonance and
the first and third planets are also near such a resonance. We find that fine-tuning in initial planet
semi-major axes, forced drift and eccentricity damping rates are required to put the system in the 8:6:4:3
chain of resonances. 

We are not investigating formation mechanisms (see \citealt{rein12} for a first investigation into this
tricky problem).

%Larger eccentricity damping lets you stabilize things as we have seen from \citet{batygin10} choice of $K$ parameter.
%As described by \citet{batygin10} and referring to Morbi, order of capture is important.

\section{N-body Simulations with a Planetesimal Disk}
Using orbital elements from the converging integration discussed above, we run N-body simulations of the
four planets, initially in resonance, in the vicinity of a planetesimal disk and about a central star.
These simulations were run with the software QYMSYM \citep{moore11}. QYMSYM is a GPU-accelerated hybrid
second order symplectic N-body integrator which permits close encounters similar to the \textit{MERCURY}
software package developed by \citet{chambers99}.

We began each simulation with four planets in a resonant chain and an external planetesimal disk
comprised of 1024 equal mass planetesimal particles. Orbital elements for the four planets in the
integration shown in Figure \ref{fig:imran} at time $t=70,000$ orbits, and listed in Table
\ref{tab:tab1}, were used as initial conditions for these integrations.

The planetesimal disk particles ranged in semi-major axis from $a_{min} = 1.95$, just outside the
outermost planet, to $a_{max} = 2.95$. The distribution of planetesimal semi-major axes is flat with
probability independent of semi-major axis within $a_{min}$ and $a_{max}$. The initial eccentricity and
inclination distributions were chosen using Rayleigh distributions with the mean eccentricity $e$
equivalent to twice the mean value of the inclination $i$ and $i = 0.01$. The initial orbital angles
(mean anomalies, longitudes of pericenter and longitudes of the ascending node) for the planetesimals
were randomly chosen. We work in units of the initial orbital period of the innermost planet.

Six simulations were run, each with different planetesimal mass. We included a test case with massless
disk particles to check the stability of the integration lacking planetesimals. We labeled the
simulations by the mass of the planetesimal disk, with simulation Z corresponding to a massless
planetesimal disk while simulations M, E, and N correspond to Mars, Earth and Neptune mass disks.
Simulation E3 and N5 refer to simulations with a third of an Earth mass and a fifth of a Neptune mass
disk, respectively. Each simulation was run for 500,000 orbital periods (or $10^4$ years as the innermost
planet's orbital period is only 7.4 days).

The timestep was chosen to be 0.08607 out of a possible $2\pi$ orbit. Given this step size, energy
conservation (measured by dE/E) was $10^{-3}$ or better for all simulations except during the simulation
labeled N (its energy conservation was $8 \times 10^{-3}$).

Due to the proximity of the planets of KOI-730 to their star, the planets practically fill their Hill
sphere. The large fraction of the Hill sphere filled limits the escape velocities of close encounters.
Our integrator does not check for collisions though it does integrate close encounters. To take into
account the large fraction of the Hill sphere filled, we adjusted the smoothing lengths during close
encounters between planets and planetesimals.

In these simulations we used a smoothing length of $s = 1 \times 10^{-3}$, which corresponds to a
distance that is slightly larger than the radii of the innermost planet (about $1.1\times$). The Hill
radius of the innermost planet, which would be the smallest of the four due to it both being the
closest to the central star as well as the least massive, is a little more than ten times larger 
(about $13.5\times$) than this smoothing length.

We have checked that the inclination distributions and extent of migration are not strongly dependent
on the assumed smoothing length. This was done by comparing the results of our simulation described
above that have planet radii sized smoothing lengths to a separate set of identical simulations only
with a smoothing length 100 times smaller.

During each simulation we computed the mass in planetesimals that crossed the orbits of the planets.
We count a planetesimal as orbit-crossing if its pericenter is less than the apocenter of the
outermost planet. The total planetesimal mass that crossed the orbits of the planets at the end of the
simulations are also listed in Table \ref{tab:tab2}. As we will discuss below, these masses can be used
to place limits on the total quantity of planetesimals that may have crossed the orbits of planets in
the KOI-730 system.

\subsection{Planetary migration in the N-body simulations}
\begin{figure}
\includegraphics[width=8.8cm]{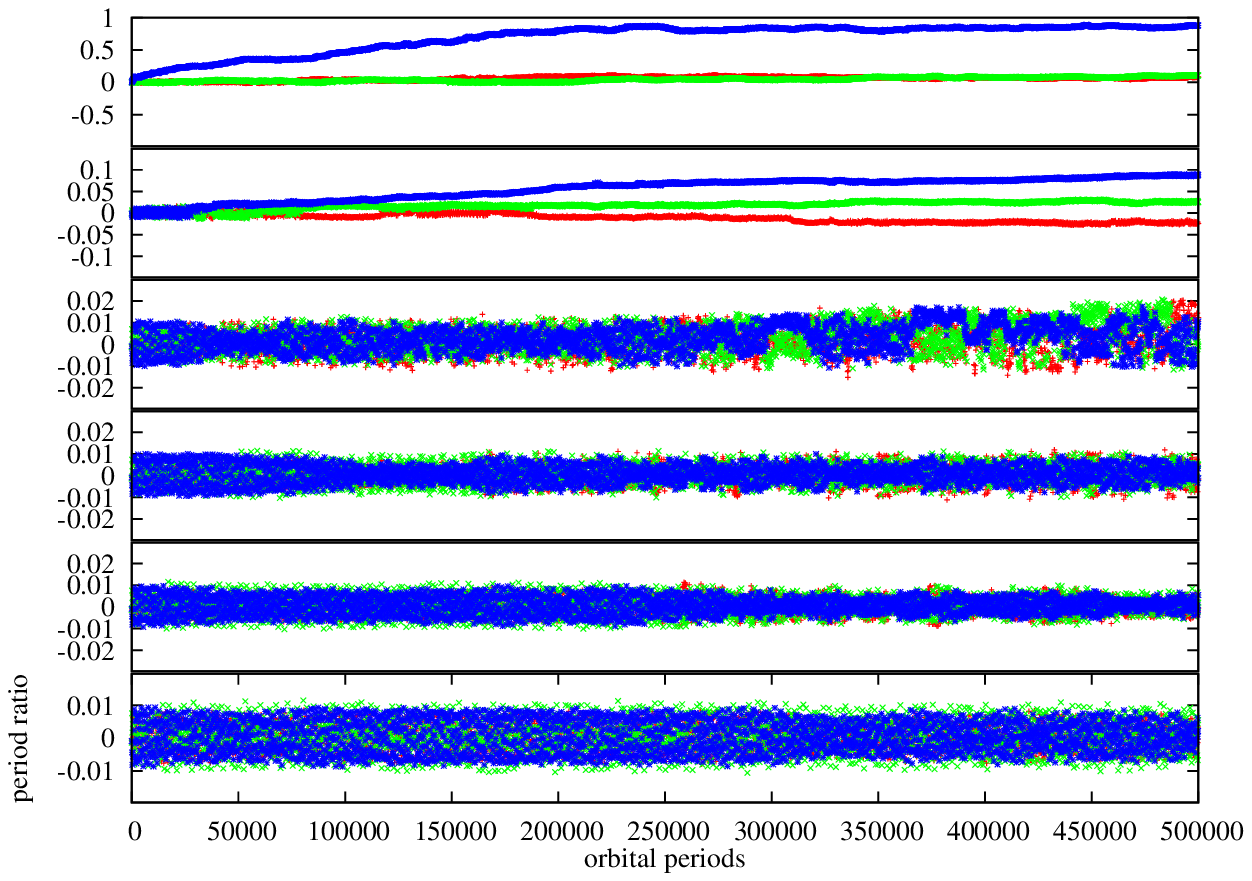}
\includegraphics[width=8.8cm]{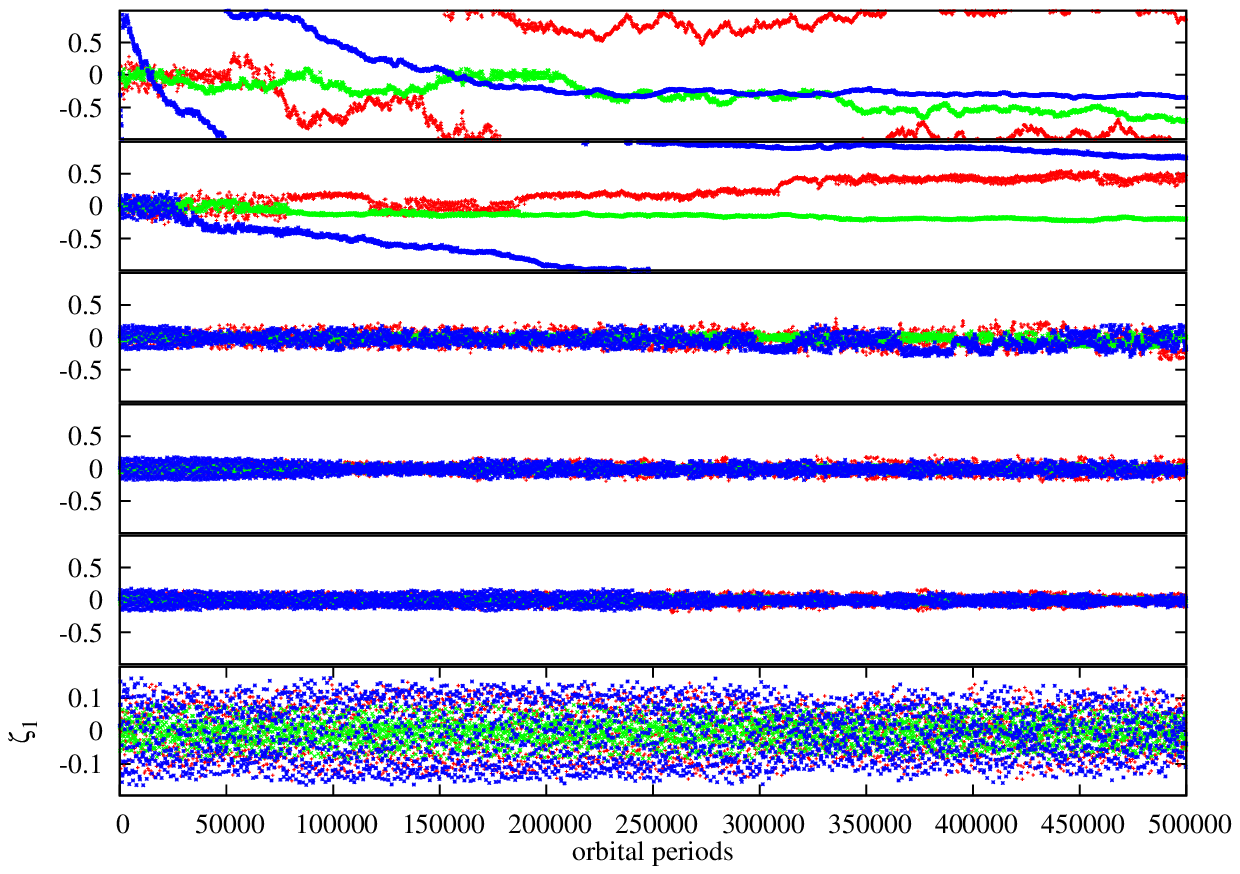}
\includegraphics[width=8.8cm]{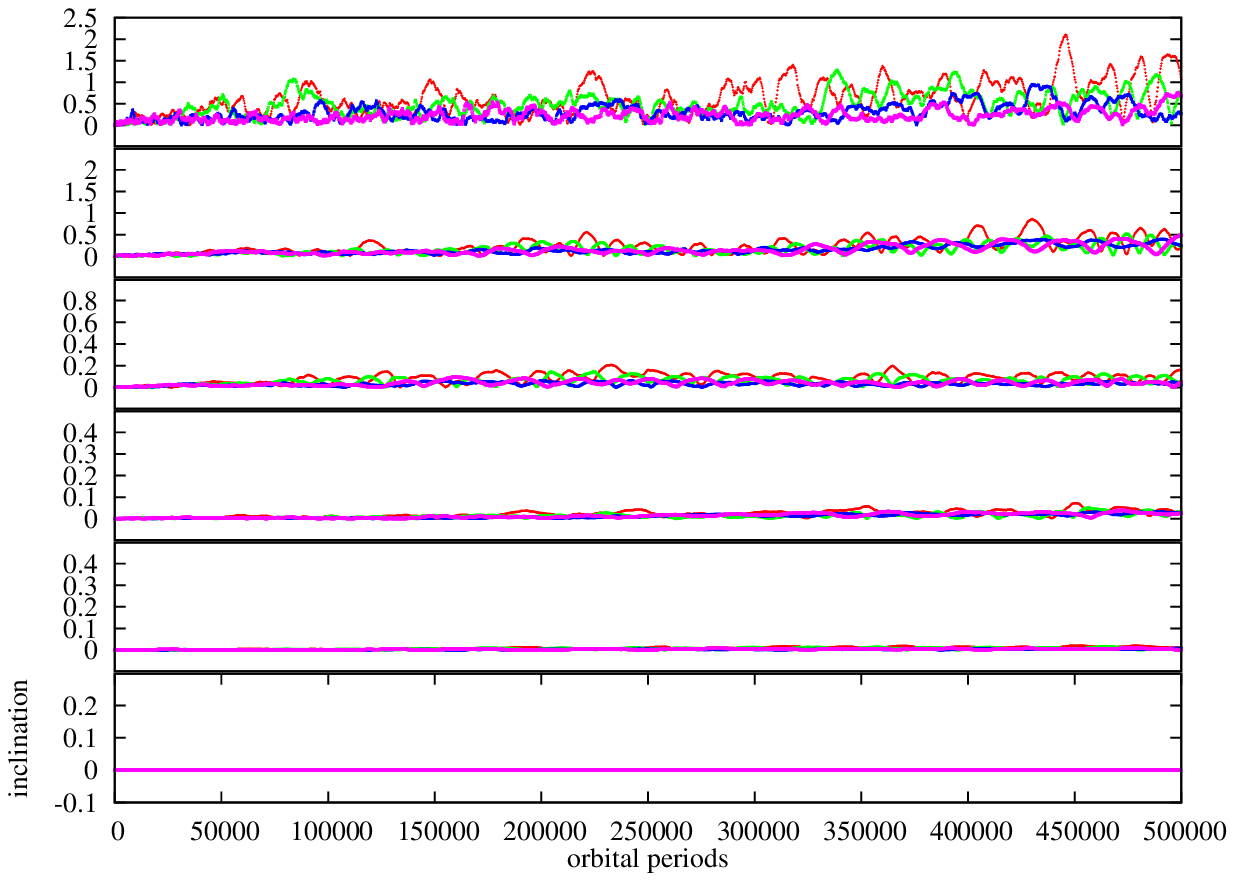}
\caption{a) Period ratios for consecutive pairs of planets subtracted by their initial value and as a
function of time. Each panel shows period ratios for a different simulation. From top to bottom, the
planetesimal disk mass decreases. Red, green and blue lines refer to the inner, middle and outer
consecutive planet pairs, respectively.
b) $\zeta_{1,1}$ parameters computed from consecutive planet period ratios and for the same simulations.
The colors are the same pairs as in a).
c) Planet inclinations in degrees as a function of time for the same simulations.
The red, green, blue and pink lines refer to the first (inner), second, third and fourth (outer)
planets, respectively.}
\label{fig:pr}
\end{figure}

Figure \ref{fig:pr}a) shows period ratios for each consecutive pair of planets as a function of time
for all 6 N-body simulations. For the inner and outer planet pairs (initially in 4:3 resonance) the
period ratio is shown subtracted by 4/3. The middle planet pair (initially in 3:2 resonance) is plotted
subtracted by 3/2. Our test case Z simulation (bottom panel in Figure \ref{fig:pr}a) remains stable
throughout the integration, consequently we are confident that the orbital elements chosen from our
capture integration are stable. It is possible that this system becomes unstable on a timescale longer
than 500,000 orbits.

The M, E3, E, N5, and N simulations contain disks with increasing mass. Because of the proximity of the
outermost planet to the inner edge of the planetesimal disk, the outer planet migrates outwards as
planetesimals cross the orbits of the planets and exchange angular momentum with them. The outer planet
migrates furthest in the simulations with highest disk mass. For the most part, period ratios increase,
though as the planets separate, mean motion resonances, as they are crossed, can cause jumps in both
eccentricity and semi-major axis. At some times we see a signature of three-body
resonances, when the period ratio for one consecutive pair (of three planets) increases as the period
ratio of the other consecutive pair decreases \citep{quillen11}.

The Z simulation remains locked in resonance throughout the integration. However, we see that the M and
E3 simulations no longer maintain their chains of MMR's at 250k and 175k orbits respectively. The E
simulation is removed from resonance at approximately 50k orbits. This allows us to estimate the mass
in planet-crossing planetesimals required to pull this system out of resonance for each simulation,
finding on average $m_c \sim M_{Mars}$. The N and N5 simulations have their planets out of resonance
very quickly due to the larger amount of orbit crossing mass early in the simulation. In both of those
simulations the planets are out of resonance too quickly for us to accurately determine how much mass
was orbit crossing. Simulations E, E3, and M all have total crossing masses within an order of magnitude
of each other.

For each integration we compute $\zeta_{1,1}$ (equation \ref{eqn:zeta}) as a function of time for each
consecutive pair of planets. Figure \ref{fig:pr}b shows the $\zeta_{1,1}$ function computed from all
consecutive planet pairs for the same simulations. Variations in the $\zeta_{1,1}$ parameter are largest
for the most massive disk. We see from Figure \ref{fig:pr}b that $\zeta_{1,1}$ approaches -0.2 at a time
of about 50k orbits for planets 3 and 4. This allows us to estimate the amount of planetesimal disk mass
that would move a system sufficiently out of resonance to contribute to the peak seen in the $\zeta_1$
distribution of the Kepler multiple planet candidates. Based on the mass in orbit-crossing planetesimals
in simulation N5 we estimate $m_c \approx ~ M_\oplus$ is required to pull a system initially in a
resonant chain sufficiently far apart to give a $\zeta_1$ equivalent to the position of the peak seen in
the $\zeta_1$ distribution by \citet{fabrycky12} (see their Figure 5).

We note in Figure \ref{fig:pr}b that we find that the $\zeta_{1,1}$ function does not always decrease for each
planet pair. Furthermore, during migration we see a range of $\zeta_{1,1}$ values.  To produce a peak in
the distribution of $\zeta_1$ we would require a very specific or fine tuned value for the mass in
orbit-crossing planetesimals for multiple planet systems that are originally in resonant chains.

Figure \ref{fig:pr}c shows the inclinations of all planets as a function of time. We see that the planet
inclinations increase to a couple of degrees in the simulations with the most massive disks. We have
checked that simulated planetesimals have not been ejected at high velocity. The masses of our simulated
planetesimals is quite low (see Table \ref{tab:tab1}) and the inclinations slowly increase. The increases
in planet inclination are unlikely to be numerically generated during encounters and due to extreme
scattering events. The smooth increase in inclinations can be attributed to a combination of gravitational
heating caused by scattering of planetesimals and to crossing of vertical resonances resulting from the
migration of the planets (as seen by \citealt{libert11}). 

Figure \ref{fig:pr}a implies that planet inclinations in closely spaced multiple planet systems depend 
on the total planetesimal mass that has crossed the orbits of the planets.  

\citet{fabrycky12} find that mutual inclinations for the multiple planet systems lie in the range 
$1.0 - 2.3^\circ$ and that their distribution is well modeled by a Rayleigh distribution with standard
deviation $\sqrt{\langle i^2 \rangle} =  1.8^\circ$. We see in Figure \ref{fig:pr}c that the inclinations
rise above this value when the total mass in orbit crossing planetesimals exceeds $17 M_\oplus$, or about
one Neptune mass. Therefore, closely packed systems similar to the KOI-730 system likely have not
experienced an era similar to the Late-Heavy Bombardment in our Solar system. From this comparison we
tentatively place a limit on the total mass in orbit crossing planetesimals during the lifetime
of the KOI-730 system of $m_c \la  M_\oplus$.

\section{Discussion and Conclusion}
Using an N-body integrator with the addition of drag causing convergent migration and eccentricity damping,
we constructed a resonant chain for four planets with period ratios 8:6:4:3, and used planet masses estimated
for the KOI-730 multiple planet system. Orbital elements from this integration were then used as initial
conditions for N-body simulations with the four planets which included an external planetesimal disk.
Interactions with the planetesimal disk allowed the planets to migrate, primarily diverging rather than
converging, as expected.

We find that one Earth mass of orbit-crossing planetesimals is sufficient to pull a system similar to
the KOI-730 system out of its chain of mean motion resonances. As planet radii estimated from transit data
depend on stellar radii, it is possible that we have over or underestimated the masses of the planets in the
KOI-730 system. The distance migrated by a planet should scale with the mass in planetesimals that it
interacts with. However, mean motion resonant widths are larger when planet masses are larger. We find that
it is more difficult to form a resonant chain via resonant capture and such chains are less stable when the
planets are more massive. Consequently, the amount of material required to pull a system out of resonance
may not scale linearly with planet mass. Nevertheless, we expect that if the true planet masses are larger
than adopted here, a larger mass in orbit-crossing planetesimals would be required to pull this system out
of resonance and vice-versa if the planets are less massive.

After the system is out of resonance, we find that it remains stable. Our simulations do not exhibit
planet/planet orbit crossing events. The KOI-730 system, comprised of sub-Neptunian mass planets, can be
compared to the HR8799 system, comprised of hyper-Jovian mass planets in resonance. When the HR8799 system
is pulled out of resonance, the system is extremely unstable \citep{god09,moore12}.

The KOI-730 system is likely in (or very close to) a resonant chain. Because interactions with planetesimals
and tidal interactions between planets primarily cause planetary orbits to slowly diverge
\citep{papa11,lithwick12} and so pull systems out of resonance, we can be fairly confident that processes
following depletion of a gas disk did not put this system in resonance. Because the system is currently
near or in resonance, only a small mass in planetesimals could have ever crossed the orbits of these planets.
We infer that this system either lacks a debris disk or contains one that is so diffuse or stable that less
than an Earth mass of debris has ever crossed the orbits of these planets.

We also find that an Earth mass of orbit-crossing planetesimals can cause the planets to migrate far enough
that the system lies sufficiently outside of resonance to resemble the Kepler systems with $\zeta_1 \sim -0.2$
where the peak of the distribution lies \citep{fabrycky12,liss11}. However we expect that different planetary
systems would have different quantities of orbit-crossing planetesimals. Consequently we would not expect that
a narrow peak in the $\zeta$ distribution would arise in a distribution of planetary systems. Fine tuning in
the quantity of planetesimals may be required to account for the peak seen in the $\zeta$ distribution. The
tidal damping scenario for pulling pairs of planets away from resonance \citep{lithwick12,papa11} may more
naturally account for a peak in $\zeta$.
% doesn't work for multiple planet systems 

We find that the more an initially flat planetary system interacts with planetesimals, the higher the mean
planet inclinations. If somewhere between a few earth masses to about a Neptune mass of planetesimals cross
the orbits of the planets, the planet inclinations can increase to a few degrees. This is sufficiently high
that they would not be all simultaneously be detected in transit. The tidal circularization scenario would
not give a relation between migration distance and inclination. However over short distances, as planets
migrate and they cross vertical resonances, we would expect a correlation between migration distance (and
so distance out of resonance) and inclination. Study of the relation between the inclination and period
distributions of the Kepler systems may differentiate between roles of tidal forces and planetesimals. 
% in affecting xxxx.

It is possible that the transiting multiple planet systems discovered by the Kepler mission are more compact
or lower mass than radial velocity discovered planetary systems. Can we differentiate between the tidal
interaction mechanism for pulling systems out of resonance and that caused by interactions with planetesimals?  
If planetesimals interact with planets, then it is likely that planet inclinations can increase as planets
cross vertical resonances. It may be possible to differentiate between these two mechanisms based on
inclination distributions as tidal interactions likely do not increase inclinations but planet/planetesimal
interactions can. Note that \citet{libert11} have shown that resonant capture for higher planet mass systems
can also induce inclination variations. However, additional mechanisms, such as turbulence associated with a
gas disk or secular perturbations from distant planets could also affect the inclination distributions. Future
studies can probe the role of planetesimals, and migration associated with scattering them, in accounting for
the inclination distributions of the Kepler planetary systems.

We have focused here on interactions with a low eccentricity planetesimal disk. When it encounters a planet,
a high eccentricity planetesimal is less strongly gravitationally focused than a low eccentricity one.
Consequently, high eccentricity objects are less effective at scattering a planet or inducing migration.
Our limit on the total mass in orbital crossing planetesimals can be considered a lower limit as we began
with a low eccentricity disk. Future studies can explore the possibility that compact Kepler systems harbor
massive outer planetary systems and high eccentricity cometary populations.

In summary, we believe that closely spaced, low inclination multiple planet Kepler systems likely have either
low mass or extremely stable debris disks. There appears to be a relation between the inclination and amount
of migration for a planet. The inclination distributions may make it possible to differentiate between dynamical
scenarios for pulling planets out of resonance. Due to the improbability of a Late Heavy Bombardment like
scenario for KOI-730, we believe these inclinations are probably caused by crossing vertical resonances.

\vskip 0.3 truein

Acknowledgments. This work was in part supported by NSF through award AST-0907841. We thank Hal Levison for
helpful discussion.

%STUFF TO MENTION:
%Mention that we tried the drift code with non-zero inclinations (of
%order 1 degree) and no differences seen
%in ae behavior other than expected secular inclination perturbations
%
%Also getting $1^\circ$ inclination
%Comparing HARPS and Kepler surveys: The alignment of multiple-planet systems
%
%gravitational heating should depend on the coulomb log (max to min
%scale) and so should
%not be strongly dependent on the smoothing length assumed.


\begin{thebibliography}{}
		
\bibitem[Batalha et al.(2012)]{batalha12}
Batalha, N. M., et al. 2012, in press, arXiv1202.5852

%EARLY DYNAMICAL EVOLUTION OF THE SOLAR SYSTEM: PINNING DOWN THE INITIAL CONDITIONS OF THE NICE MODEL
\bibitem[Batygin \& Brown(2010)]{batygin10}
 Batygin, K. \& Brown, M. E. 2010, ApJ, 716, 1323%-1331

%Planetary migration and extrasolar planets in the 2/1 mean-motion resonance
\bibitem[Beaug\'e et al.(2006)]{beauge06}
Beauge, C.,  Michtchenko , T. A., \& Ferraz-Mello, S. 2006, MNRAS, 365, 1160%-1170

\bibitem[Chambers(1999)]{chambers99}
Chambers J. E., 1999, MNRAS, 304, 793

\bibitem[Fabrycky \& Murray-Clay(2010)]{fabrycky10}
Fabrycky D.C., \& Murray-Clay R.A., 2010, ApJ, 710, 1408

%Architecture of Kepler's Multi-transiting Systems: II. New investigations with twice as many candidates
\bibitem[Fabrycky et al.(2012)]{fabrycky12}
Fabrycky, D. et al. 2012, ApJS in press, http://arxiv.org/abs/1202.6328

%Evolution of Migrating Planet Pairs in Resonance
\bibitem[Ferraz-Mello et al.(2003)]{ferraz03}  
Ferraz-Mello, S.,  Beaug\'e, C., \& Michtchenko, T. A. 2003, CeMDA, 87, 99

\bibitem[Figueira et al.(2012)]{figuera12}
Figueira, P., Marmier, M., Boue, G., Lovis, C., Santos, N. C., Montalto, M., Udry, S., Pepe, F., \& Mayor, M.
2012, A\&A in press, arXiv.1202.2801

\bibitem[G\'ozdziewski \& Migaszewski(2009)]{god09}
G\'ozdziewski K., \& Migaszewski C., 2009, MNRAS, 397, 16

%Kepler-9: A System of Multiple Planets Transiting a Sun-Like Star, Confirmed by Timing Variations
\bibitem[Holman et al.(2010)]{holman10}
Holman, M. et al. 2010, Science , 330 , 51

%Evolution of planetary systems in resonance
\bibitem[Kley et al.(2004)]{kley04}
Kley, W., Peitz, J., \& Bryden, G. 2004, A\&A, 414, 735	

%Dynamics and Origin of the 2:1 Orbital Resonances of the GJ 876 Planets
\bibitem[Lee \& Peale(2002)]{lee02}
Lee, M. H., \& Peale, S. J.	 2002, ApJ, 567, 596	

%Diversity and Origin of 2:1 Orbital Resonances in Extrasolar Planetary Systems
\bibitem[Lee(2004)]{lee04}	
Lee, M. H. 2004, ApJ, 611, 517	

%Late Orbital Instabilities in the Outer Planets Induced by Interaction with a Self-gravitating Planetesimal Disk
\bibitem[Levison et al.(2011)]{levison11}
Levison, H. F., Morbidelli, A., Tsiganis, K., Nesvorny, D., \& Gomes, R. 2011, AJ, 142, 152		
	
%Architecture and Dynamics of Kepler's Candidate Multiple Transiting Planet Systems
\bibitem[Lissauer et al.(2011)]{liss11}
Lissauer, J. J. et al. 2011, ApJS, 197, 8

%Almost All of Kepler's Multiple Planet Candidates are Planets
\bibitem[Lissauer et al.(2012)]{liss12}
Lissauer, J. J. et al. 2012, ApJ, in press,  arXiv1201.5424

%Trapping in three-planet resonances during gas-driven migration
\bibitem[Libert  \& Tsiganis(2011)]{libert11}
Libert, A.-S.,  \& Tsiganis, K. 2011, Celest Mech Dyn Astr, 111, 201%-218

%Resonant Repulsion of Kepler Planet Pairs
\bibitem[Lithwick \& Wu(2012)]{lithwick12}
Lithwick, Y., \& Yanqin Wu, Y.  2012, in press, arxiv1204.2555

%On a Hermite integrator with Ahmad-Cohen scheme for gravitational many-body problems
\bibitem[Makino \& Aarseth(1992)]{makino92}
Makino, J., \& Aarseth, S. J. 1992, PASJ, 44, 141

\bibitem[Matsumura et al.(2010)]{matsumura10}
Matsumura, S., Thommes, E. W., Chatterjee, S., \& Rasio, F. A. 2010, ApJ, 714, 194

\bibitem[Moeckel \& Armitage(2012)]{moeckel12}
Moeckel, N. \& Armitage, P. J. 2012, MNRAS, 419, 366

\bibitem[Moore \& Quillen(2011)]{moore11}
Moore A., \& Quillen A., 2011, NewAST, 16, 445

\bibitem[Moore \& Quillen(2012)]{moore12}
Moore, A., \& Quillen, A. C. 2012, in prep

%Dynamics of the Giant Planets of the Solar System in the Gaseous Protoplanetary Disk and Their Relationship to the Current Orbital Architecture
\bibitem[Morbidelli et al.(2007)]{morbi07}
Morbidelli, A., Tsiganis, K., Crida, A., Levison, H. F., \& Gomes, R. 2007, AJ, 134, 1790	

%On the migration-induced resonances in a system of two planets with masses in the Earth mass range
\bibitem[Papaloizou \& Szuszkiewicz(2005)]{papa05}
Papaloizou, J. C. B., \& Szuszkiewicz, E. 2005, MNRAS, 363, 153

%Tidal interactions in multi-planet systems
\bibitem[Papaloizou(2011)]{papa11}
Papaloizou, J. C. B. 2011, CeMDA, 111, 83

%On the dynamics of resonant super-Earths in disks with turbulence driven by stochastic forcing
\bibitem[Pierens et al.(2011)]{pierens11}	
Pierens, A., Baruteau, C., \& Hersant, F. 2011, A\&A, 531, 5

%reducing the probability of capture into resonance
\bibitem[Quillen(2006)]{quillen06}
Quillen, A. C.  2006, MNRAS, 365, 1367

% three body resonances
\bibitem[Quillen(2011)]{quillen11}
Quillen, A. C. 2011, MNRAS, 418, 1043

%Planet-Planet Scattering in Planetesimal Disks
\bibitem[Raymond et al.(2009]{raymond09}		
Raymond, S. N., Armitage, P. J., \& Gorelick, N.  ApJ, 669, 88
%Planetesimal disks cause damping of e, i and for large planets put in resonant chains sometimes

%Traditional formation scenarios fail to explain 4:3 mean motion resonances
\bibitem[Rein et al.(2012)]{rein12}
Rein, H., Payne, M. J., Veras, D., \& Ford,  E. B. 2012, MNRAS in press, arXiv:1204.0974

\bibitem[Thommes et al.(2008)]{thommes08}
Thommes, E. W., Bryden, G., Wu, Y., \& Rasio, F. A.  2008, ApJ, 675, 1538	
%From Mean Motion Resonances to Scattered Planets: Producing the Solar System, Eccentric Exoplanets, and Late Heavy Bombardments

\bibitem[Tsiganis et al.(2005)]{nice}
Tsiganis, K., Gomes, R., Morbidelli, A., \& Levison, H. F. 2005, Nature, 435, 459
%, Origin of the orbital architecture of the giant planets of the Solar System

%Identifying non-resonant Kepler planetary systems
\bibitem[Veras \& Ford(2012)]{veras12}
Veras, D., \& Ford, E. B.	 2012, MNRAS, 420, L23
	
 %A Survey of Multiple Planet Systems
%\bibitem[Wright(2010)]{wright10}     
%Wright, J. T. 2010,  European Astronomical Society Publications Series,  42, 3%-17
%EAS....42....3W  
%Proceedings of Extrasolar Planets in Multi-Body Systems: Theory and Observations, Extrasolar Planets in Multi-Body Systems, Toru?, Poland.
\bibitem[Wisdom(1980)]{wisdom80}
Wisdom, J. 1980, AJ, 85, 1122

\bibitem[Wright et al.(2009)]{wright09}   
Wright, J. T., Upadhyay, S., Marcy, G. W., et al. 2009, ApJ, 693, 1084

\bibitem[Wright et al.(2011)]{wright11}  
Wright, J. T., Fakhouri, O., Marcy, G. W., et al. 2011, PASP, 123, 412
	
\end{thebibliography}
\end{document}